\documentclass[12pt]{iopart}
\usepackage[dvips]{graphicx}

\begin{document}
\title{Exciton self-trapping in bulk polyethylene}
\author{D. Ceresoli\dag, M. C. Righi\ddag, E. Tosatti\dag\S, S. Scandolo\S, 
  G. Santoro\dag\ and S. Serra$\sharp$}

\address{\dag\ International School for Advanced Studies (SISSA) and 
DEMOCRITOS National Simulation Center, Trieste, Italy.}
\address{\ddag\ Dipartimento di Fisica, Universit\`a di Modena, Modena, Italy.}
\address{\S\ International Center for Theoretical Physics (ICTP), Trieste, Italy.}
\address{$\sharp$\ Pirelli Labs - Materials Innovation, V.le Sarca 222, I-20126, Milano, Italy.}

\begin{abstract}
We studied theoretically the behavior of an injected electron-hole pair in 
crystalline polyethylene. Time-dependent adiabatic evolution by ab-initio 
molecular dynamics simulations show that the pair will become self-trapped 
in the perfect crystal, with a trapping energy of about 0.38 eV, with 
formation of a pair of trans-gauche conformational defects, three C$_2$H$_4$ 
units apart on the same chain. The electron is confined in the inter-chain 
pocket created by a local, 120$^\circ$ rotation of the chain between the 
two defects, while the hole resides on the chain and is much less bound. 
Despite the large energy stored in the trapped excitation, there does not 
appear to be a direct non-radiative channel for electron-hole recombination. 
This suggests that intrinsic self-trapping of electron-hole pairs inside
the ideal quasi-crystalline fraction of PE might not be directly relevant 
for electrical damage in high-voltage cables.
\end{abstract}
\maketitle

\section{Introduction}
Thanks to its large band-gap ($\sim 8.8$ eV), to its chemical inertness,
and to its easy processing, polyethylene (PE) is the material of choice 
in extruded electric cables for high-voltage applications. 
The highly insulating properties of PE however do not prevent a tiny flow 
of charge carriers, both electrons and holes, injected at the very high 
electric fields realized under normal high voltage operation conditions. 
Understanding and controlling this flow is of course extremely important 
for improving the cable performances. For example, dielectric breakdown 
depends crucially on the mobility of the injected charge carriers. 
Aging and eventual failure of the cable is also believed to be connected 
to local damage induced by charge flow and/or recombination\cite{dissado}. 
From a physicist's perspective, it is of primary importance to understand 
the basic electronic properties of PE which underlie the observed 
phenomenology. In particular it is crucial to explore the detailed 
nature of the electronic states corresponding to the highest valence-band 
state (holes), to the lowest conduction band states (electrons), and to 
the simultaneous presence of holes and electrons whose bound state we 
will refer to with the standard name of excitons\cite{bassani}. 

Excitons are particularly interesting in this context, as it is 
known~\cite{laurent} that under high field {\em and} polarity reversals,
electroluminescence --  field-induced emission of photons -- is observed 
with particular intensity, plausibly due to radiative electron-hole 
recombination. However electroluminescence is also known to open the 
way to electric damage. When in fact the electron-hole recombination 
energy, amounting to several eV, is released in a non-radiative channel,
it is suddenly turned into ion kinetic energy, and will most likely
fuel important local structural and eventual electric damage. The possible 
existence of channels for this kind of damage is as yet unexplored.

Recently, considerable progress was made in the understanding of the 
nature of, separately, electron and hole states in PE. Photoemission 
studies~\cite{valence} agree well with electronic structure 
calculations~\cite{miao,jones,serra:interchain} in showing that the filled 
valence bands are made up of C$-$C (and of C$-$H) bonding states, 
propagating very effectively along the polymeric chain, but rather 
ineffectively between the chains. We shall refer to states of this kind, 
that are heavily chain-localized, as \emph{intrachain} states. The 
lowest empty conduction band states were found on the contrary 
(so far only theoretically) to propagate quite well in all directions in
three 
dimensional space, and also to display a maximum amplitude no longer on 
the chains, but rather in the empty space between the chains. Such a state 
was referred to as an \emph{interchain} state~\cite{serra:interchain,cubero}.
The interchain nature of the conduction states is held responsible for 
some uncommon physical properties such as the \emph{negative electron affinity} or NEA~\cite{righi:nea}, characteristic of PE. 

An extra electron injected in a soft insulator like bulk PE will 
not remain indefinitely free, delocalized, and conduction band-like. As 
in other soft materials, the extra electron can be expected to cause some 
kind of spontaneous static distortion of the lattice -- a small polaron -- 
whose effect will be to trap the electron itself in the volume encompassed 
by the deformation (\emph{self-trapping}). Detailed 
calculations~\cite{serra:selftrapping} indicated the existence of such a 
polaron state in bulk crystalline PE. It consists of a short PE chain 
segment -- limited by a pair of \emph{trans-gauche} conformational defects 
-- which undergoes a large rotational distortion, with the extra electron 
locally bound, and self-trapped. No self-trapping, conversely, was found 
for holes, owing presumably to the stiffness of the C$-$C and C$-$H covalent 
bonds in the chains onto which the hole intra-chain wave function is 
localized.

The scope of this work is to address the issue of how a bound electron-hole
pair -- an exciton -- will behave in this respect once injected in PE. 
Electron-hole pairs can be created through electrical injection of hot 
carriers. Although ordinary carrier mobility in real PE is extremely small, 
of order $10^{-10}$ cm$^2$/V, some electrons and holes do nonetheless get 
injected at opposite electrodes when the field overcomes the threshold 
for space charge formation. Once inside, electrons and holes drift toward 
one another. 
Due to their (screened) Coulomb attraction they will as soon as possible 
merge to form bound exciton pairs. Excitons can of course also be created 
optically, \cite{bassani1}  by absorption of UV photons of energy higher 
than the PE energy gap. 

These considerations and the related questions call for an investigation 
of the electron-hole pair (\emph{exciton}) state in crystalline PE, in 
particular of its accompanying structural deformation, and of its possible 
self-trapped state in the perfect material. We approached this problem 
theoretically, making use as in our previous electron and hole studies, of 
constrained
density functional electronic structure calculations, that possess enough 
quantitative accuracy to be predictive.

We will present material in the following order. First we shall describe in 
Section II the methods and computational techniques used. In Section III we 
will study the exciton in perfect crystalline PE, and demonstrate its 
self-trapping, the associated distortion similar to some extent but not 
identical to that of the isolated electron.  Section IV will be devoted to 
discussion of the possible consequences, and to our conclusions.

\section{Computational techniques}
We carried out first-principle electronic structure calculations of PE,
using density-functional methods~\cite{DFT}, based on 
plane-wave expansion and pseudopotentials. Technical details are identical
to those given in previous papers~\cite{serra:dynamical,serra:interchain,
serra:selftrapping}. It should be stressed that the choice of a large plane 
wave basis set in these calculations is particularly important, in order to 
capture the best possible real nature of conduction states. Restricted local 
basis sets, otherwise very effective for describing chemical bonding and 
well localized valence states, should be considered with suspicion, 
as they may easily fail to describe properly the extremely extended, plane 
wave-like {\it interlayer} conduction states~\cite{serra:interchain}. 
Calculations were carried out within 
the BLYP gradient corrected energy functional~\cite{BLYP}. The ion-electron 
interactions were described by norm-conserving Martins-Troullier 
pseudo\-pot\-entials~\cite{MTpseudo}. Wave functions were expanded in plane 
waves, with a plane wave energy cutoff of 40 Ry. Structural optimization and 
molecular dynamics (MD) simulations were carried out with a steepest descent
and a  Car-Parrinello algorithm~\cite{CarParrinello}, respectively. 
The time step 
for the Car-Parrinello dynamics was set to 5 atomic units (0.6 fs) and the 
electronic mass to 200 atomic units. 
Ab initio, finite temperature MD simulations where carried out with the 
help of a Nos\'e thermostat with a fictitious mass corresponding to a 
frequency of 40 THz.

The van der Waals attraction between the chains -- a numerically small but 
qualitatively important contribution to the stability of PE which is not 
accounted for by quasi-local approximations to the DFT functional such as BLYP
-- was included empirically through an extra two body interatomic potential 
with a parametrized $r^{-6}$ tail of the form:
\begin{equation}\label{eq:vdw}
  V_{vdw} = - \frac{1}{2} \sum_{i,j} \frac{C_{i,j}}{r^6_{i,j}}f_c(r_{i,j}),
\end{equation}
where $f_c$ is a function cutting off at short distance. All the parameters 
entering (\ref{eq:vdw}) were given the same values as in Serra 
\emph{et al.}~\cite{serra:dynamical}, optimized to correctly reproduce the 
equilibrium structure, stability, elastic constants and thermal properties of 
neutral PE. This ensures that we begin with a sound overall physical 
description of PE, prior to the excitation, consisting of the simultaneous 
introduction of an electron and a hole.

A first basic question of the present excited-state calculations is how in 
fact to introduce the electron-hole pair. A natural choice would be to 
introduce a triplet exciton. Theoretically, a triplet exciton is the $S=1$ 
ground state, and  can in principle be introduced in a calculation by 
forcing a total spin of one in a spin-polarized calculation\cite{marco}. 
Experimentally, triplet excitons could be created by impact energy loss 
exchange processes of low energy electrons,~\cite{bocchetta} whereby an 
incoming spin down electron will fall in energy to occupy an empty conduction 
state, while kicking away a spin up electron, and thus creating a spin up 
valence hole. Although its actual energy location and lifetime are presently 
unknown, a long-lived {\em triplet} $S=1$ exciton must surely exist in PE, as in 
all other molecular solids, with an excitation energy below the ordinary, 
singlet excitation gap. Practical and computational limitations force us
however to abandon 
this costly spin-polarized procedure, and to consider a simpler alternative 
solution. We mimic here the presence of an exciton by simply constraining an
occupancy of
one, instead of two, for the highest pair of occupied Kohn-Sham eigenvalues 
in a non-spin-polarized calculation (constrained DFT).
Spin-contamination problems (related
to the fact that such an exciton is, in a non-polarized calculation, neither a 
singlet nor a triplet pure state \cite{spincontamination}) are expected to 
be minor in this case. In fact the hole and the electron states, one 
intrachain and the other interchain, overlap poorly in space, making the 
contribution of electron-hole exchange relatively small, and negligible with 
respect to the large exciton creation energy of many eV (approximately
equal to the band gap of about 8.8 eV). This kind of approximation will of course not hold any more 
when, as {\em e.g.,} in aromatics, both electron and hole states share the 
same electronic nature; or else when, even in impure PE, electron and hole 
may both become tightly localized onto some chemical defects. However, in the 
present case of defect free PE we believe that this represents a highly 
reasonable approximation. It can also be noted that the probable slight 
overestimate of excitation energy relative to the true triplet state will 
cancel at least a fraction of the well-known DFT energy gap deficit, thus 
representing a small improvement, rather than a deficiency. 

We carried out first of all the electronic structure calculations and the 
structural optimization in this model exciton state. 
A sophisticated many-body description (GW-BS~\cite{louie})
of the electron-hole interaction would be desirable to describe the excitonic 
energies, but the associated structural relaxations
can be obtained quite accurately within a constrained DFT approach, as
reported for the self-trapped exciton in a conjugated organic 
polymer~\cite{artacho}.
Subsequently, starting from this state we also carried out a reasonable number 
of Car-Parrinello molecular dynamics steps, in order to explore effectively 
the potential energy surface of the system.

\section{Results}
Perfect crystalline PE possesses a base-centered orthorhombic crystal 
structure with lattice parameters $a=$ 4.93 \AA, $b=$ 7.4 \AA\ and $c=$ 2.534 
\AA~\cite{crystalstructure}. The unit cell contains two polymer chains, 
running parallel to the $c$ direction. The two chains are rotated by 
$\pm$ 42$^\circ$ around the $c$ axis, and the CH$_2$ units are in a 
trans-planar conformation, namely the carbon skeleton forms a zig-zag chain 
lying entirely on a plane (figure 1 of ref.~\cite{serra:interchain}).

The calculated DFT/BLYP energy band gap for ground state PE is 6.46 
eV~\cite{serra:dynamical}, in perfect agreement with
previous calculations~\cite{valence,jones} but
somewhat smaller than the experimental gap of about 8.8 eV.
Due to that, the energy available to create a distortion, in our
approximation, will represent a lower bound to that in real PE.
We modeled the system with four parallel chains, each seven -CH$_2$-CH$_2$- 
units
long in a periodically repeated simulation cell. We first performed a 
calculation starting from the perfect crystalline PE configuration. However, 
the self-trapped state of the excess electron is known to involve a structural 
rearrangement of the chains~\cite{serra:selftrapping}, and this rearrangement 
might be hindered by energy barriers when starting from ideal perfect 
crystalline PE. To circumvent that, we also performed a series of calculations 
starting from distorted atomic configurations morphologically similar to that 
of the self-trapped electron. In particular, we created two gauche defects on 
the same chain at a distance of one, two, three, and four C$_2$H$_4$ units, 
the chain segment between the two defects rotated by 120$^\circ$ with respect 
to the crystalline PE structure. 

\begin{figure}\begin{center}
  \includegraphics[width=7.5cm,clip]{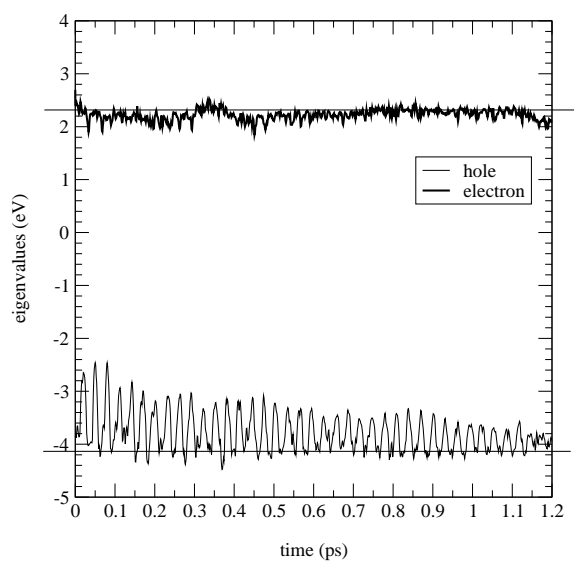}
  \caption{Kohn-Sham (KS) eigenvalues of crystalline PE as a function of time,
  during a constant temperature molecular dynamics run.
  The two horizontal lines represent the band edges.}
  \label{fig:crystal_eigen}
\end{center}\end{figure}

We start by describing first the results obtained for the undistorted 
structure. We relaxed first the structure by steepest descent, both in the 
ground state and in the excited state. In both cases, deviations from the 
trans-planar conformation were negligible. The ground state HOMO-LUMO gap of 
6.46 eV is only slightly reduced after excitation to 6.36 eV: the hole 
Kohn-Sham (KS) eigenvalue 0.09 eV above the top of the valence band, and the 
electron KS eigenvalue 0.01 eV below the bottom of the conduction band. 
The resulting reduction of the energy gap is 0.1 eV. Such a small value
may be due to the well known gap-problem of LDA.
We did not attempt to
perform GW-BS~\cite{louie} calculations in order to compute the exciton
binding energy.
Starting from this relaxed static structure, we performed next a molecular 
dynamics simulation at 300 K. The atomic positions oscillated about the 
trans-planar conformation and no defects were spontaneously created. The 
behavior with simulation time of the electron and hole Kohn-Sham eigenvalues 
is shown in fig.~\ref{fig:crystal_eigen}: the hole eigenvalue oscillates with 
a magnitude of about 0.5 eV, but the gap remains substantially unchanged for 
the duration of the simulation (1.5 ps), showing no evidence for self trapping, 
at least in this relatively short time.  

\begin{figure}\begin{center}
  \includegraphics[height=7.5cm,angle=90]{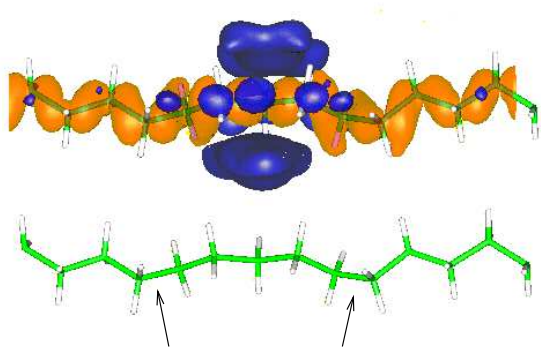}
  \caption{Iso-density surface of the self-trapped exciton; hole-state in 
  light gray; electron state in dark gray.
  The bare polymeric chain is shown on the left for sake of clarity.
  The two small arrows indicate the position of the gauche defects.}
  \label{fig:3units_iso}
\end{center}\end{figure}

\begin{figure}\begin{center}
  \includegraphics[width=7.5cm,clip]{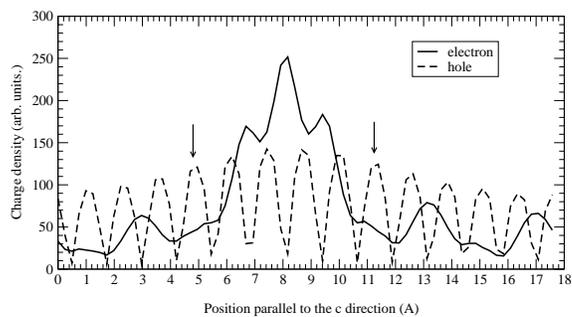}
  \caption{Charge density profile parallel to the $c$ direction.
  Small arrows indicate the position of the gauche defects.}
  \label{fig:3units_charge}
\end{center}\end{figure}

Similar results were found in the simulations which started from the distorted 
structures. Annealing the system by molecular dynamics at room temperature for 
0.3 ps led to states where the distortion persisted, suggesting that each 
distortion is a local minimum of the energy landscape, and that different 
minima are separated by energy barriers, that will take a long time to 
cross. In order to determine the energetically favored structure, we optimized 
by steepest-descent the atomic positions of each distorted structure with 
$n=1,...,4$ rotated C$_2$H$_4$ units. The total energy calculated at the local 
minimum of each distorted structure is reported in Table I. We found in this 
way that the self-trapped exciton with lowest energy consisted of 
\emph{three} 
rotated C$_2$H$_4$ units (fig.~\ref{fig:3units_iso}). 
Comparison with the self-trapped electron case 
(without hole), where the distortion consisted of seven rotated units, 
supports the notion that the electron-hole attraction contributes substantially 
to the electron localization, reducing the self-trapped polaron size from 
seven for the electron to three for the electron-hole pair. While in the 
electron-only case the singly occupied electron level sank about 0.5 eV
below the conduction band bottom, that level (upper singly occupied level) is 
found here to lie a mere 0.04 eV below the conduction band, 
most likely on account of the extra kinetic energy required by localization 
in the smaller 3-unit pocket~\cite{serra:selftrapping}.
The hole energy level (lower singly occupied 
level) is 0.01 eV above the top of the valence band, corresponding to a much 
weaker localization of the hole wavefunction in the neighborhood of the 
trapped and strongly localized electron, as seen in 
fig.~\ref{fig:3units_charge}. The remainder of the total energy gain, is thus
to be ascribed to the screened electron-hole Coulomb attraction.
Fig.~\ref{fig:3units_charge} shows the electron and hole charge-density 
profile parallel to the $c$ direction; the dashed line is the square modulus 
of the hole wave-function; the maxima correspond to the center of the the 
C$-$C bonds. The solid line is the square modulus of the electron 
wave-function; the position of the gauche defects is indicated by the two 
short arrows. The hole is only slightly affected by the presence of the 
electron but tends to be more localized in the proximity of the electron, due 
to Coulomb attraction.

Fig.~\ref{fig:3units_iso} shows the iso-density surfaces of the hole and 
electron states together with the polymeric chain. The electron state is 
clearly \emph{inter-chain} and is trapped in the volume encompassed by the 
deformation.

\section{Discussion and conclusions} 
Our calculations indicate that \emph{self-trapped electron-hole pairs} 
should exist in crystalline polyethylene, with a distortion 
pattern qualitatively 
similar to -- although quantitatively different from -- that of the 
self-trapped electron. The necessary energy to detrap 
the exciton
(consisting of reduction of band-gap, increase of kinetic energy due to
confinement, creation of two gauche defects, rotation of a segment of chain
and electron-hole Coulomb interaction)
is about 0.3--0.4 eV. The exciton state, although very high in 
absolute energy, is apparently long-lived,
displaying no apparent direct channel for 
non-radiative recombination. This suggests that exciton self-trapping in ideal 
quasi-crystalline PE might not be directly relevant for electrical damage 
and implies that release of the energy (several eV) stored in the electron-hole
pair should take place in other forms. Likely candidates for that would seem 
conformational or chemical defects, possibly via a combination of non-radiative 
and radiative decay processes. Further studies will be needed to explore such 
more complex scenarios.  

\ack
We thank Dr. G. Perego (Pirelli Cavi) for very fruitful discussions.
Work in SISSA was sponsored by MIUR FIRB RBAU017S8R004,
FIRB RBAU01LX5H, and COFIN 2003 and COFIN 2004, 
as well as by INFM Progetto Calcolo Parallelo.



\begin{table}\begin{center}
  \begin{tabular}{c c}
  \hline \hline
  Number of rotated & Energy difference respect \\
  C$_2$H$_4$ units    & to the undistorted case (eV) \\
  \hline
    1 & 1.22 \\
    2 & 1.55 \\
    {\bf 3} & {\bf -0.38} \\
    4 & -0.05 \\
  \hline \hline
  \end{tabular}
  \caption{Self-trapped exciton: energy difference respect to the 
     undistorted case.}
  \label{tab1}
\end{center}\end{table}

\end{document}